\documentclass[reprint,amsmath,amssymb,aps,prl,superscriptaddress]{revtex4-1}
\usepackage[dvipdfmx]{graphicx}
\usepackage{bm}

\begin{document}
\title{
Incommensurate Short-Range Order
in $S = 1$ Triangular Lattice Ising Antiferromagnet
}
\author{Yusuke Tomita}
\email{ytomita@shibaura-it.ac.jp}
\affiliation{College of Engineering, Shibaura Institute of Technology, Saitama 330-8570, Japan}
\author{Milan \v{Z}ukovi\v{c}}
\affiliation{Department of Theoretical Physics and Astrophysics, Faculty of Science, P. J. \v{S}af\'arik University, Park Angelinum 9,041 54 Ko\v{s}ice, Slovakia}

\date{\today}

\begin{abstract}
We study the $S = 1$ triangular lattice Ising antiferromagnet by
Monte Carlo simulations. Frustrations between a major
antiferromagnetic third-neighbor interaction $J_3$ and
a minor ferromagnetic nearest-neighbor interaction $J_1$
cause incommensurate short-range orders at intermediate temperatures.
At low temperatures (below $T/J_3 \lesssim 0.2$ for $J_1/J_3 = -1/3$),
the system exhibits fourfold periodic ordered state.
In the short-range order phase, the system shows a glassy two-step relaxation.
We demonstrate that the features of the short-range order
are attributed to the cooperation between the frustrations
and the nonmagnetic Ising spin states which is a particular feature of
the integer spin systems.
\end{abstract}

\pacs{
05.50.+q, 
64.60.De, 
64.70.Q-, 
75.50.Mm  
}
\keywords{geometrical frustration; Ising antiferromagnet; triangular lattice; Monte Carlo simulation}

\maketitle


Suppressed magnetic order in geometrically frustrated systems\cite{Ramirez94}
has been attracting attention in material science and statistical physics.
In particular, understanding disordered states solely by frustration
(i.e., without randomness) has been an important challenge in both theoretical
and experimental physics, because such disordered states can be qualitatively
different from high-temperature paramagnetic states in terms of topological
order\cite{Wen90,Kitaev06,Isakov11} or pseudo-critical behavior\cite{Lin14}.
A simplest frustrated spin model is the triangular lattice
antiferromagnetic Ising model\cite{Wannier50}.
Strong frustration in the model suppresses the spontaneous emergence of order
at finite temperatures.
An extension of the model to the $S = 1$ Ising model\cite{ZukovicBobak13}
alleviates the frustration, and the extended model
exhibits the Berezinskii-Kosterlitz-Thouless (BKT) transitions\cite{Berezinskii71,KosterlitzThouless73}.
Thus, in view of realizing a cooperative short-range order phase at
finite temperature, frustration in the former model is too strong
while it is too weak in the latter.
In this Letter, we demonstrate that incommensurability introduced
in the $S = 1$ model by adding the further-neighbor interaction
can stabilize a peculiar short-range ordered phase with
a glass-like \textit{dynamical} property.


The Hamiltonian of the $J_1$-$J_3$ model is given by
\begin{equation}
\mathcal{H} = J_1\sum_{\langle i, j\rangle}\sigma_i\sigma_j
+ J_3\sum_{[i, k]}\sigma_i\sigma_k,
\label{eq:Hamiltonian}
\end{equation}
where $\sigma_i(\in \{0, \pm 1\})$ represents an $S = 1$ Ising spin at site $i$.
In the right hand side of Eq.~(\ref{eq:Hamiltonian}),
the first term represents the ferromagnetic nearest-neighbor
interactions ($J_1 < 0$), and the second term represents
the antiferromagnetic third-neighbor interactions ($J_3 > 0$).
For $J_1 = 0$, the model comprises four decoupled sublattices of
the $S = 1$ triangular lattice Ising model.
Unlike the $S = 1/2$ triangular lattice Ising antiferromagnetic system,
the antiferromagnetic interactions give rise to
the BKT transitions
at finite temperatures\cite{ZukovicBobak13, Landau83}.
The BKT transition of the model is described by the Z$_6$ symmetry breaking
corresponding to the three-sublattice structure ($+1, 0, -1$):
Since the triangular lattice is a tripartite lattice,
each of the three states of the $S = 1$ Ising spin can avoid
neighboring the same state.
Such favorable states in energetical point of view degenerate
into six states.
The ordered state is almost the same as the 120$^\circ$ structure
in the six-state antiferromagnetic clock model on the triangular lattice,
but there are inhomogeneities in energies at sites.
By introducing a small nearest-neighbor ferromagnetic coupling $J_1$,
the four triangular sublattices are coupled, leading to a different
ground state. Here, we note that $|J_1| \ll J_3$ is not necessarily
an unrealistic assumption. Such a situation can be realized when the
system possesses some combination of direct and superexchange couplings
as in a $S = 1$ quantum antiferromagnet NiGa$_2$S$_4$.
Below we set $J_1/J_3 = -1/3$ corresponding to an estimate in
this compound\cite{Nakatsuji05}.
By analyzing the Fourier transform of the exchange couplings,
we find that a magnetic instability may occur at incommensurate (IC)
wave vectors $\pm(k, 0)$, $\pm(k/2, \sqrt{3}k/2)$,
and $\pm(k/2, -\sqrt{3}k/2)$ for $-4 < J_1/J_3 < 0$,
where $k$ is determined by\cite{TamuraKawashima08,*TamuraKawashima11}
\begin{equation}
J_1/J_3 = - [2(\sin k + \sin 2k)]/[\sin k + \sin(k/2)].
\label{eq:wavenumber}
\end{equation}
Using Eq.~(\ref{eq:wavenumber}), we obtained $k \simeq 1.922$
for $J_1/J_3 = -1/3$.

To investigate thermal equilibrium properties of the $J_1$-$J_3$
model, Monte Carlo (MC) simulations are executed.
The Suwa-Todo algorithm\cite{SuwaTodo10} that minimizes the average of
rejection rate is employed to improve spin updates.
The triangular lattice of the system size $L$ contains
$N (=L^2)$ spins, and the periodic boundary conditions are imposed.
To estimate thermal averages,
$N_{\textrm{MCS}} = 2\times 10^7$ ($1\times 10^7$) MC steps are
executed for $L =$ 144, 192, and 288 ($L =$ 48, 72, and 96).
The simulations start from the highest temperature ($T/J_3 = 1.1$),
and the temperature is gradually lowered with the step $\Delta T/J_3 = 0.01$.
For thermalization at each temperature, $0.2\times N_{\textrm{MCS}}$ MC steps
are discarded before measuring observables.
To estimate statistical errors, independent 10 samples are simulated.
Hereafter, Boltzmann constant $k_{\text{B}}$ is set to unity.

\begin{figure}[h]
  \includegraphics[width=.4\textwidth]{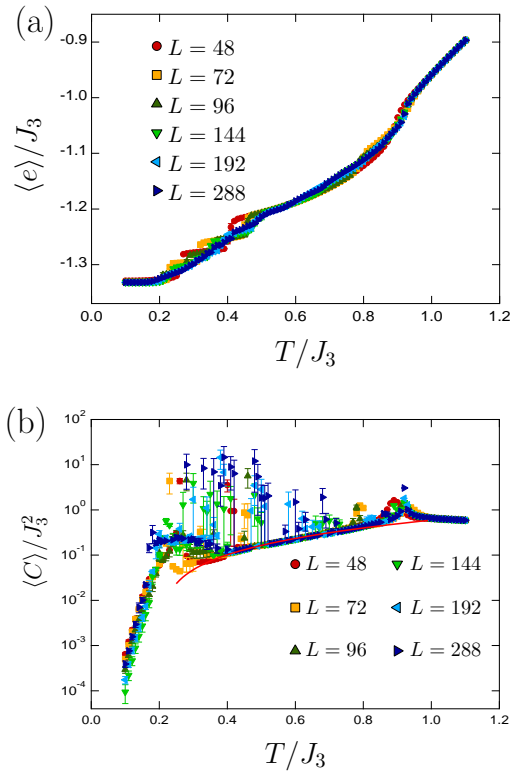}
  \caption{\label{fig:energy}
(Color online)
(a) Plot of the internal energy per site.
Error bars are smaller than size of marks.
(b) Semi-logarithmic plot of the specific heat.
The solid curve is a quadratic curve that fits the intermediate data.
Since the specific heat jump at $T/J_3 \sim 0.2$
is extremely large ($10^5$ order)
for $L = 192$ and 288, they are omitted for the visibility.
}
\end{figure}

Figures~\ref{fig:energy} show the internal energy
and specific heat per site for several system sizes.
We observed the specific heat jump 
at $T/J_3 \sim 0.2$ for $L = 192$ and 288
(data are not shown for the visibility).
Scatterings of the specific heat data are observed
in $0.2 \lesssim T/J_3 \lesssim 0.9$.
To analyze the origin of the data scattering, 
we calculate the structure factor,
\begin{equation}
S(\bm{k}) = \langle \sigma_{\bm{k}}\sigma_{-\bm{k}}\rangle,
\label{eq:sfactor}
\end{equation}
where $\sigma_{\bm{k}}$ denotes the Fourier transform of the spins
and $\bm{k}$ is the wavevector.

\begin{figure}[!h]
  \includegraphics[width=.4\textwidth]{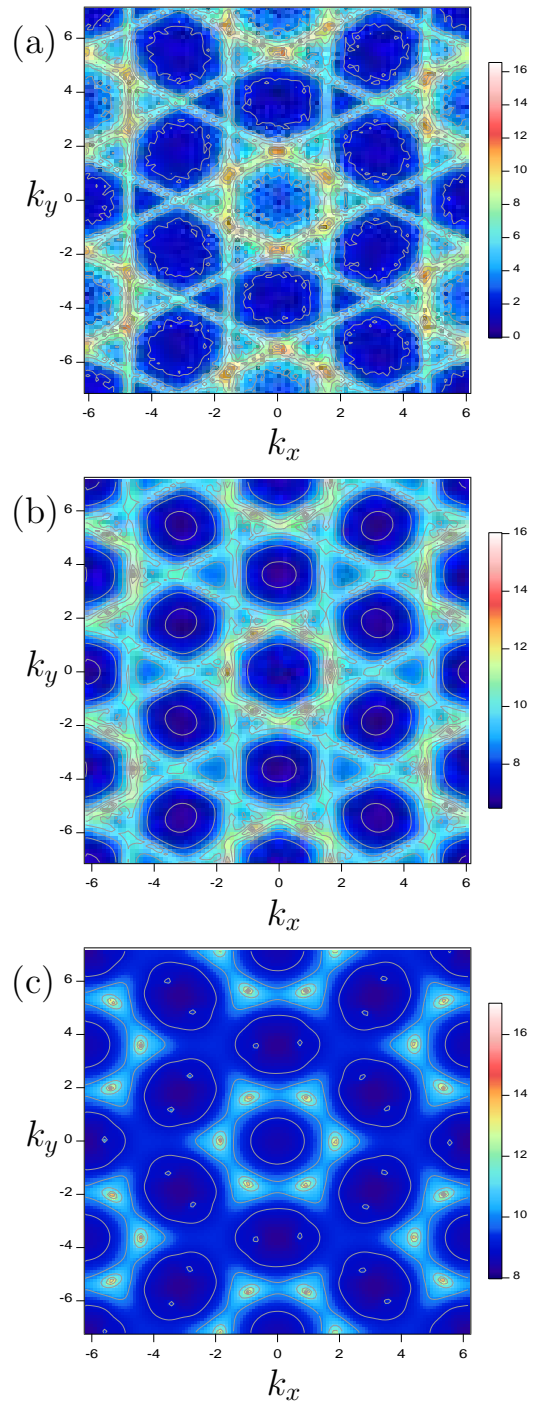}
  \caption{\label{fig:sfactor}
(Color online)
Structure factors for $L = 192$ at three different temperatures are plotted.
Temperatures of the plots are (a) $T/J_3 = 0.15$,
(b) $T/J_3 = 0.3$, and (c) $T/J_3 = 0.7$.
To improve visualization,
the natural logarithmic scale is adopted for structure factors.
Contour lines are superimposed.
}
\end{figure}

Structure factors for $L = 192$ are plotted in
Figs.~\ref{fig:sfactor}(a)-~\ref{fig:sfactor}(c).
The structure factor at $T/J_3 = 0.7$ shows six peaks.
The positions of the peaks are the same positions
given by Eq.~(\ref{eq:wavenumber}) within the accuracy of
the lattice discreteness.
This coincidence is explained by the universality class of the six-state
clock model. The clock model has two transition points,
and the universality class of the transition point at the higher temperature
is the same as the XY model.
Since thermal fluctuations screen the discreteness of the clock model,
the system exhibits the universality class of the XY model.
In the $J_1$-$J_3$ model, each triangle of Ising spins behaves
like a planer spin variable\cite{ZukovicBobak13, Landau83}
around $T/J_3 = 0.7$,
so that the structure factor exhibits the development of
the spin-spin correlations same as the continuous spins.
Hereafter we call the magnetic short-range order (SRO) as the planer order.

The opposite situation arises at low temperatures ($T/J_3 \lesssim 0.2$).
In the low temperatures, thermal fluctuations are too weak to screen
the discreteness of spin variables, and an ordered state resulting
from the discreteness of the Ising spin emerges.
The ordered state is a fourfold periodic spin configuration
($\uparrow\uparrow\downarrow\downarrow$).
Hereafter we call the magnetic structure as the low-$T$ order.
Since spin configurations obtained by MC simulations
consist of domains of the fourfold periodic pattern,
six peaks are observed at
$\pm$(0, $\pi/\sqrt{3}$), $\pm$($\pi/2$, $\pi/(2\sqrt{3})$), and
$\pm$($\pi/2$, -$\pi/(2\sqrt{3})$) in the structure factor.
The perfect fourfold structure gives the internal energy per site
$\langle e\rangle/J_3 = -4/3$,
whereas internal energies obtained by the MC simulations
are slightly higher than $-4/3$.

In a rather wide window of the intermediate temperatures 
($0.2 \lesssim T/J_3 \lesssim 0.9$), 
the peaks in the structure factor become less sharp
(i.e., the correlation length decreases) as lowering the temperature,
while the peak positions remain almost the same as those at $T/J_3 = 0.7$.
In Fig. 3, we show the correlation length defined by
\begin{equation}
\xi = \frac{1}{2\sin(\Delta k/2)}
\sqrt{\frac{S(k_{\text{max}})}{S(k'_{\text{max}})} - 1},
\label{eq:xi}
\end{equation}
where $\Delta k = 2\pi/L$ and $S(k_{\text{max}})$ and $S(k'_{\text{max}})$ are,
respectively, the structure factor averaged at six peaks and
that averaged at neighboring sites of the peaks.
At least according to snapshots of the spin configuration (not shown),
the temperature dependence of the magnetic domain is as follows:
Planer SRO domains adjoin each other
at rather high temperatures ($0.5 \lesssim T/J_3 \lesssim 0.9$),
while the domains are separated by thick domain walls
at lower temperature.
At rather low temperatures ($0.2 \lesssim T/J_3 \lesssim 0.5$),
domain walls become thicker as the temperature is lowered.
The low-$T$ order appears in the thick domain walls,
and it becomes the dominant order at $T/J_3 \sim 0.2$.

\begin{figure}[h]
  \includegraphics[width=.4\textwidth]{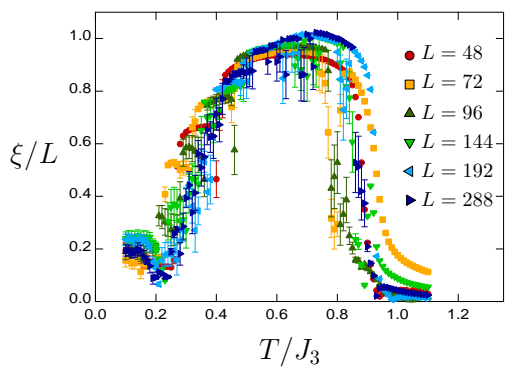}
  \caption{\label{fig:xi}
(Color online)
Plot of the normalized correlation length $\xi/L$ for
several system sizes.
}
\end{figure}

The competition between the planer IC order and the low-$T$ fourfold
periodic order reminds us the modulated and floating IC phase
in the axial next-nearest neighbor Ising (ANNNI) model\cite{Selke88}.
The modulated phase consists of a lot of periodic phases,
and each phase is separated by the first-order transition.
Though the magnetic structure in the $J_1$-$J_3$ model is incommensurate, 
there might be temperatures that the IC magnetic patterns approximately
match the lattice spacing.
At such temperatures, energy fluctuations become larger,
and data scattering in the specific heat is observed.

\begin{figure}[h]
  \includegraphics[width=.4\textwidth]{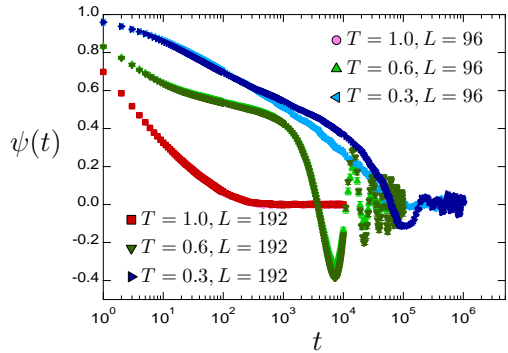}
  \caption{\label{fig:psi}
(Color online)
Plot of the time-displaced correlation function
for $L = 96$ and $L = 192$ at three different temperatures.
}
\end{figure}

If the above analogy between the $J_1$-$J_3$ model and the ANNNI model
is appropriate, the magnetic domains exhibit unusual relaxation process.
The most distinguished dynamical feature of the ANNNI model is
the floating IC phase, in which floating (not locked) domain walls
are observed.
To investigate the dynamical feature of the $J_1$-$J_3$ model,
we calculated a time-displaced correlation function,
\begin{equation}
\psi(t) = \frac{\langle \sum_i^N\sigma_i(t)\sigma_i(0)\rangle}
{\langle \sum_i^N\sigma_i^2(0)\rangle},
\end{equation}
where $\langle\cdots\rangle$ denotes the thermal average.
Since this quantity is affected by spin update methods,
we changed the method from the Suwa-Todo algorithm to the heat bath method.
In Fig.~\ref{fig:psi}, we plot the correlation functions.
The correlation function shows exponential decays
in paramagnetic phase ($T/J_3 = 1.0$),
whereas the correlations in the planer IC phase
($T/J_3 = 0.6$ and $0.3$) gradually decay in time.
Oscillating behavior in $\psi$'s in the planer IC phase
indicates that domains are floating,
since displaced domains make $\psi$ negative.
Qualitative differences in relaxations of the correlation function
between that of $T/J_3 = 0.6$ and of $T/J_3 = 0.3$ are quite interesting.
At the first stage ($t \lesssim 10$), $\psi$'s at $T/J_3 = 0.6$ show
fast decays, while $\psi$'s at $T/J_3 = 0.3$ show monotonic decays.
The fast decay is probably attributed to relaxations around
nonmagnetic states ($\sigma = 0$).
Since interaction energies around nonmagnetic states are zero regardless of
their spin states, relaxations should be faster
than those around magnetic states.
The heterogeneities in the energy and dynamics remind us
of the rattling motion in glassy materials\cite{MuranakaHiwatari95,Kob97,Weeks00}.
The nonmagnetic states give rise to \textit{cage structures} as in glass formers
and kinetically constrained models\cite{FredricksonAndersen84, KobAndersen93,BiroliMezard02}.
As the temperature is lowered, correlations of the low-$T$ order
are enhanced, so that the density of nonmagnetic spins at $T/J_3 = 0.3$
is not sufficient enough to show the rattling motion.
At the second stage ($10 \lesssim t \lesssim 10^3$),
$\psi$'s at $T/J_3 = 0.6$ show quite slow relaxation;
slopes are flatter than those at $T/J_3 = 0.3$.
This slow relaxation is explained by the crossover
from the BKT phase.
As we mentioned above, each triangle of Ising spins
is regarded as planer spin variable when thermal fluctuations
are sufficiently strong.
Considering that the peak positions of the structure factor
in the planer SRO phase are located near those of planer spin systems,
the triple Ising spins are behaving like planer spins.
In addition, the sharpness of the peaks indicates that
systems at higher temperatures are more likely to be affected by 
the BKT phase.
The plateau regime observed at $T/J_3 = 0.6$ is similar to
the $\beta$-relaxation in glassy materials.
To the best of our knowledge, the mechanism of the emergence
of the $\beta$-relaxation in glassy materials has not been unveiled.
However, in the $J_1$-$J_3$ model, the emergence of the plateau regime
can be interpreted as the crossover from the BKT phase.

One might be interested in constructing a spin model which
exhibits no long-range order at finite temperatures.
Such a spin model is probably realized by introducing chemical potential
term $\mu\sum_i\sigma^2_i$ into the $J_1$-$J_3$ model.
By setting the parameter $\mu$ sufficiently positively large,
the planer SRO phase pushes out the low-$T$ order phase.


Albeit many differences, thermal properties of the Ising $J_1$-$J_3$ model
are surprisingly quite similar to those of NiGa$_2$S$_4$\cite{Nakatsuji05}
at least at a phenomenological level.
Though our model has the low-$T$ order phase,
the ordered phase is probably eliminated by introducing
the chemical potential term as mentioned above.
By including such a chemical potential term,
the $J_1$-$J_3$ model would exhibit no long-range order at finite temperatures,
which is consistent with the observation in NiGa$_2$S$_4$
at least phenomenologically.
Dynamics of the $J_1$-$J_3$ model also have similarities to NiGa$_2$S$_4$.
As shown in Fig.~\ref{fig:psi}, 
the $J_1$-$J_3$ model exhibits quasi-critical relaxation
in the planer SRO phase.
Since the quasi-critical relaxation originates from the crossover
from the BKT transition, planer SRO states at higher temperatures
are more affected by the crossover.
The dynamical crossover in the $J_1$-$J_3$ model
is similar to
the signal lost in Ga-nuclear quadrupole resonance\cite{Takeya08}
and the non-monotonic linewidth growth
in electron spin resonance\cite{Yamaguchi08}.

There are two grounds for the similarities.
The first one is that the $J_1$-$J_3$ model possesses the emergent
planer spin symmetry in the planer SRO phase, which might be related
to the easy-plane anisotropy of the $S = 1$ Heisenberg spins
in NiGa$_2$S$_4$\cite{Yamaguchi10}.
The other similarity is the one between $S = 1$ Ising and quantum Heisenberg
spins in a sense that the Hilbert space is identical.
The discreteness of the Ising spin is consistent with
the discreteness of the quantum spin state.
A notorious deficiency of classical Heisenberg spin systems
is that they estimate the specific heat at unity in the limit of $T = 0$.
Therefore, classical continuous spin systems are not suitable
for studying of quantum spin systems at low temperatures.
Moreover, the $S = 1$ Ising spin emulates the nonmagnetic state ($\sigma = 0$)
in quantum spin systems.
The classical Heisenberg spin does not have this feature,
and the lack of the feature causes the long-range order
at low temperatures\cite{TamuraKawashima08,TamuraKawashima11,Stoudenmire09}.
Even though the $S = 1$ quantum Heisenberg $J_1$-$J_3$ model exhibits
an antiferromagnetic long-range order when $J_1/J_3 = -1/3$,
short-range order emerges in the range of $-4.0 \lesssim J_1/J_3 \lesssim -2.2$\cite{Rubin12}.
The importance of the feature has already been pointed out
by one of the authors in the study of variable-length
Heisenberg spin model\cite{TomitaKawashima11}.
The significant influence by nonmagnetic states
is also consistent with experimental observations.
Nambu and Nakatsuji studied the spin-size dependent impurity effects
by Ni$_{1-x}$A$_x$Ga$_2$S$_4$
[A = Zn ($S = 0$), Fe ($S = 2$), Co ($S = 3/2$), and Mn ($S = 5/2$)]\cite{NambuNakatsuji08,*NambuNakatsuji11}.
They found that the integer spin impurities
do not bring about a drastic change, 
whereas the canonical spin glass-like behavior
is observed with the substitution of the magnetic impurities
with half-odd integer spins.
Understanding the difference between integer and half-integer spins
is easy if the analogy between 
the $S = 1$ Ising $J_1$-$J_3$ model and NiGa$_2$S$_4$
is applicable.
Due to pinning of floating planer IC domains at
half-integer spin impurity sites,
the \textit{floating} IC phase is taken over
by the \textit{randomly locked} IC phase.
On the other hand, integer spin impurities
are not able to pin a floating domain,
so that no significant change emerges.
Though the symmetry of the Ising spin is different from
the Heisenberg spin, the Ising spin system seems to be more suitable
for describing the low temperature properties of NiGa$_2$S$_4$
than the classical Heisenberg spin models.

\textit{Conclusions}.---
We have performed MC simulations of the 2D $S = 1$ Ising $J_1$-$J_3$ model.
The model exhibits three phases;
the paramagnetic, the intermediate planer SRO, and
the fourfold periodic ordered phase.
In the planer SRO phase, the true long-range order is absent,
since magnetic domains are floating.
However, the time-displaced correlation function shows quasi-critical
relaxation, because of the crossover from the BKT transition.
The two-step relaxation process observed at higher temperatures
is quite similar to the relaxation process in glass formers.
Quite interestingly, the relaxation process is
emerged from purely geometric frustration in the simple spin model.
We expect that
our model provides a good foothold for understanding slow dynamics
in frustrated and/or glassy systems.
While the symmetry of our spin model is different from
that of NiGa$_2$S$_4$,
because of the identical structure of the Hilbert space,
we show that our spin model can phenomenologically reproduce
several aspects observed in NiGa$_2$S$_4$.
Even though our model is not the exact model of NiGa$_2$S$_4$,
studying the model will help to understand frustrated magnets.

We thank Y. Kamiya for valuable discussions
and useful comments on the manuscript.
The computation in the present work was performed on computers
at the Supercomputer Center, Institute for Solid State Physics,
University of Tokyo.

\bibliography{tomita_zukovic}

\end{document}